\documentclass[prd,preprintnumbers,twocolumn,eqsecnum,floatfix,a4paper,nofootinbib,superscriptaddress]{revtex4}
\usepackage{color}
\usepackage{calc}
\usepackage{amsmath,amssymb,graphicx}
\usepackage{amssymb,amsmath}
\usepackage{bm}
\usepackage{microtype}
\usepackage{booktabs}
\usepackage{times}
\usepackage{subfigure}
\usepackage[varg]{txfonts}
\usepackage[utf8]{inputenc}
\definecolor{LinkColor}{rgb}{0.75, 0, 0}
\definecolor{CiteColor}{rgb}{0, 0.5, 0.5}
\definecolor{UrlColor}{rgb}{0, 0, 0.75}
\allowdisplaybreaks
\DeclareFontFamily{OT1}{pzc}{}
\DeclareFontShape{OT1}{pzc}{m}{it}{<-> s * [1.10] pzcmi7t}{}
\DeclareMathAlphabet{\mathpzc}{OT1}{pzc}{m}{it}
\usepackage[breaklinks, plainpages=false, colorlinks=true, anchorcolor=cyan, linkcolor=red, citecolor=blue, urlcolor=magenta, bookmarks=false]{hyperref}
\usepackage{natbib}
\usepackage{blindtext}
\begin{document}
\renewcommand{\thefigure}{\arabic{figure}}
\setcounter{figure}{0}
\def\I{{\rm i}}
\def\E{{\rm e}}
\def\D{{\rm d}}

%%%%%%%%%%%%%%%%%%%%%%%%%%%%%%%%%%%%%%%%%%%%%%%%%%%%%%%%%%%%%%%%%%%%%%%%%%%%%%%%%%%%%%%%%%%%%%%%%%%%%%%%%%%%%%%%%%%%%%%%%%%%%%%%%%%%%%%%%%%%%%%%%%%%%%%%%%%%%%%%%%%%%%%%%%%%%%%%%%%%%%%%%%%%%%%%%%%%%%%%%%%%%%%%%%%%%%%%%%%%%%%%%%%%%%%%%%%%%%%%%%%%%%%%%%%%%%%%%%%%%%%%%%%%%%%%%%%%%%%%%%%%%%%%%%%%%%%%%%%%%%%%%%%%%%%%%%%%%%%%%%%%%%%%%%%%%%%%%%%%%%%%%%%%%%%%%%%%%%%%%%%%%%%%%%%%%%%%%%%%%%%%%%%%%%%%%%%%%%%%%%%%%%%%%%%%%%%%%%%%%%%%%%
\title{Enigmatic Velocity Dispersions of Ultra-Diffuse Galaxies\\ in Light of Modified Gravity Theories and Radial Acceleration Relation}

\author{Tousif Islam}
\email[]{tousifislam24@gmail.com}
\affiliation{ Center for Scientific Computation and Visualization Research (CSCVR), University of Massachusetts (UMass) Dartmouth, Dartmouth, MA-02740, USA}

%%%%%%%%%%%%%%%%%%%%%%%%%%%%%%%%%%%%%%%%%%%%%%%%%%%%%%%%%%%%%%%%%%%%%%%%%%%%%%%%%%%%%%%%%%%%%%%%%%%%%%%%%%%%%%%%%%%%%%%%%%%%%%%%%%%%%%%%%%%%%%%%%%%%%%%%%%%%%%%%%%%%%%%%%%%%%%%%%%%%%%%%%%%%%%%%%%%%%%%%%%%%%%%%%%%%%%%%%%%%%%%%%%%%%%%%%%%%%%%%%%%%%%%%%%%%%%%%%%%%%%%%%%%%%%%%%%%%%%%%%%%%%%
\begin{abstract}
Recent observations of anomalous line-of-sight velocity dispersions of two ultra-diffuse galaxies (UDGs) provide a stringent test for modified gravity theories. While NGC 1052-DF2 exhibits an extremely low dispersion value ($\sigma \sim 7.8_{-2.2}^{+5.6}$ km/s), the reported dispersion value for NGC 1052-DF44 is quite high ($\sigma \sim 41.0 \pm 8$ km/s). For DF2, the dynamical mass is almost equal to the luminous mass suggesting the galaxy have little to no `dark matter' in $\Lambda$CDM whereas DF4 requires a dynamical mass-to-light ratio of $\sim 30$ making it to be almost entirely consists of dark matter. It has been claimed that both these galaxies, marking the extreme points in terms of the estimated dynamical mass-to-light ratio among known galaxies, would be difficult to explain in modified gravity scenarios. Extending the analysis presented in \cite{islam2019modified}, we explore the dynamics of DF2 and DF44 within the context of three popular alternative theories of gravity [Modified Newtonian Dynamics (MOND), Weyl Conformal gravity and Modified gravity (MOG)] and examine their viability against the dispersion data of DF2 and DF44. We further show that the galactic `Radial Acceleration Relation' (RAR) is consistent with DF44 dispersion data but not with DF2.
\end{abstract}
\maketitle
%%%%%%%%%%%%%%%%%%%%%%%%%%%%%%%%%%%%%%%%%%%%%%%%%%%%%%%%%%%%%%%%%%%%%%%%%%%%%%%%%%%%%%%%%%%%%%%%%%%%%%%%%%%%%%%%%%%%%%%%%%%%%%%%%%%%%%%%%%%%%%%%%%%%%%%%%%%%%%%%%%%%%%%%%%%%%%%%%%%%%%%%%%%%%%%%%%%%%%%%%%%%%%%%%%%%%%%%%%%%%%%%%%%%%%%%%%%%%%%%%%%%%%%%%%%%%%%%%%%%%%%%%%%%%%%%%%%%%%%%%%%%%%
%%%%%%%%%%%%%%%%%%%%%%%%%%%%%%%%%%%%%%%%%%%%%%%%%%%%%%%%%%%%%%%%%%%%%%%%%%%%%%%%%%%%%%%%%%%%%%%%%%%%%%%%%%%%%%%%%%%%%%%%%%%%%%%%%%%%%%%%%%%%%%%%%%%%%%%%%%%%%%%%%%%%%%%%%%%%%%%%%%%%%%%%%%%%%%%%%%%%%%%%%%%%%%%%%%%%%%%%%%%%%%%%%%%%%%%%%%%%%%%%%%%%%%%%%%%%%%%%%%%%%%%%%%%%%%%%%%%%%%%%%%%%
\section{\textbf{Introduction}} 
van Dokkum et al \cite{van2018galaxy,van2018revised,van2019second} have recently reported two ultra-diffuse galaxies (UDGs), NGC 1052-DF2 and NGC 1052-DF4, having extraordinarily low velocity dispersions of $\sigma \sim 7.8_{-2.2}^{+5.6}$ km/s. and $\sigma \sim 4.2_{-2.2}^{+4.4}$ km/s respectively. The inferred total dynamical mass of both the galaxies is $\sim 2.0-3.0 \times 10^{8} M_{\odot}$ (where $M_{\odot}$ is the solar mass)  which is almost equal to the luminous mass of these galaxies \cite{van2018galaxy,van2018revised,van2019second}. The authors therefore suggested that these two galaxies are dominated by baryons and lack dark matter. They further claimed that the absence of dark matter (or, in other words, the low velocity dispersions) would be extremely difficult to explain in any modified theory of gravity which tries to explain the observed `mass discrepancies' in galaxies and clusters through modification of the laws of gravity without invoking the existence of the apparent `dark matter'. The discovery of DF2 (and DF4) become extremely significant when they are compared with another UDG named NGC 1052-DF44 \cite{Dokkum_2016} whose morphology and mass contents are similar to DF2 (and DF4) but exhibits a relatively high velocity dispersions ($\sigma \sim 41.0 \pm 8$ km/s) indicating the galaxy to be dark matter dominated. Thus, these two UDGs represent the extreme ends of the relative baryon and dark matter contents in a particular galaxy. While, in $\Lambda$$CDM$ (general relativity with dark matter and dark energy), these galaxies might be easier to account for, their formation channel remains poorly understood \cite{nusser2019scenario,ogiya2018tidal,leigh2019collisional,silk2019ultra}. For modified gravity theories, it is not immediately clear whether they would be able to explain the anomalous velocity dispersions of both the galaxies. Thus these two galaxies provide an acid test for any modified gravity.

We also note that the dynamics of hundreds of galaxies are well known to follow a phenomenological relation called `Radial Acceleration Relation' (RAR) \cite{McGaugh} which is regarded to be an indication of the breakdown of the Newtonian gravity (and consequently of general relativity in its weak field limit) at the galactic scale. Given DF2 and DF44 have defied the current understanding of the galaxy rotation curves and the standard stellar-to-halo-mass ratio \cite{wasserman2018deficit}, it is important to look into whether the dynamics of DF2 and DF44 are consistent with RAR.
 
In this paper, we therefore present a comprehensive study investigating the ability of modified theories of gravity in explaining the anomalous velocity dispersions of the ultra-diffuse galaxies. We choose three popular alternative theories of gravity: namely, Modified Newtonian gravity (MOND) \cite{mond1,famaey2012modified}, 
Scalar-Tensor-Vector Gravity (STVG) or MOdified Gravity (MOG) \cite{mog} and Weyl Conformal gravity \cite{weyl1,weylrot5}. These three modified theories of gravity tries to explain the dynamics of galaxies and globular clusters without assuming the existence of the `exotic' dark matter. These gravity theories successfully explain the observed rotation curves of a large number of galaxies (\cite{sanders2002modified,gentile2011things} for MOND; \cite{weylrot1,weylrot2,weylrot3,weylrot4,kt2018} for Weyl gravity; \cite{moffat2013mog,moffat2015rotational,moffat2011testing} for MOG), dispersion profiles of galactic globular clusters (\cite{scarpa2003using,scarpa2004using,scarpa2007using} for MOND; \cite{islam2018globular} for Weyl gravity; \cite{moffattoth} for MOG) and the phenomenological RAR for galaxies (\cite{ghari2019radial,kt2018} for MOND; \cite{o2019radial} for Weyl gravity; \cite{green2019modified} for MOG). 

Previous studies \cite{famaey2018mond,kroupa2018does,haghi2019new} have shown that the current observation of DF2 (and DF4) are insufficient to rule out MOND scenarios while MOG is consistent with the quoted overall dispersion value in vD18a \cite{moffat2018ngc}. It has been further shown that the high velocity dispersions of DF44 is consistent with MOND expectation but at odds with MOG \cite{haghi2019star}. In our previous paper \cite{islam2019modified}, we presented a Jeans analysis approach to investigate the velocity dispersions of DF2 and DF4 in the context of four modified theories of gravity (Weyl conformal gravity, MOND, MOG and Emergent gravity along with Newtonian gravity without dark matter). We showed that the radial dispersion data of DF2 (and the overall dispersion value of DF4) are consistent with these theories of gravity either in 2$\sigma$ or 3$\sigma$ confidence level. The analysis assumed the velocity anisotropy parameter $\xi$ to be zero. In this paper, we relax this assumption and choose a realistic generalized radially varying anisotropy model for galaxies and have revisited the issue. We further extend our study to the case of DF44 and explore any possible degeneracy in our models.

The rest of the paper is organized in the following way. We begin with a brief introduction of MOND, Weyl Conformal gravity and MOG (Section \ref{sec2}). We then present the formulation of velocity dispersions in a spherically symmetric system like ultra-diffuse galaxies (Section \ref{sec3}). In Section \ref{sec4}, we investigate the modified gravity theories in light of the observed velocity dispersions of DF2. We discuss the effects of different phenomenologically chosen parameters in our fitting exercise and look for possible degenracies in models. We repeat the exercise for NGC 0152-DF44 in In Section \ref{sec5} while Section \ref{sec6} reports a preliminary study to understand whether RAR is consistent with DF2 and DF44. We finally discuss implications of the results, point out possible caveats of the analysis (Section \ref{sec7}) and conclude (Section \ref{sec8}).
%%%%%%%%%%%%%%%%%%%%%%%%%%%%%%%%%%%%%%%%%%%%%%%%%%%%%%%%%%%%%%%%%%%%%%%%%%%%%%%%%%%%%%%%%%%%%%%%%%%%%%%%%%%%%%%%%%%%%%%%%%%%%%%%%%%%%%%%%%%%%%%%%%%%%%%%%%%%%%%%%%%%%%%%%%%%%%%%%%%%%%%%%%%%%%%%%%%%%%%%%%%%%%%%%%%%%%%%%%%%%%%%%%%%%%%%%%%%%%%%%%%%%%%%%%%%%%%%%%%%%%%%%%%%%%%%%%%%%%%%%%%%
\section{Modified Gravity Theories}
\label{sec2}
In this section, we provide an executive summary of Weyl Conformal gravity, MOND and MOG. For a detailed discussions of these theories of gravity, we refer the readers to \cite{weylrot5} (for Weyl conformal graviy), \cite{mond1,famaey2012modified} (for MOND) and \cite{mog} (for MOG).
%%%%%%%%%%%%%%%%%%%%%%%%%%%%%%%%%%%%%%%%%%%%%%%%%%%%%%%%%%%%%%%%%%%%%%%%%%%%%%%%%%%%%%%%%%%%%%%%%%%%%%%%%%%%%%%%%%%%%%%%%%%%%%%%%%%%%%%%%%%%%%%%%%%%%%%%%%%%%%%%%%%%%%%%%%%%%%%%%%%%%%%%%%%
\subsection{Weyl Conformal Gravity}
Weyl conformal gravity \cite{weyl1,weylrot5} is a covariant metric theory of gravity which employs the principle of local conformal invariance of the space-time. The action remains invariant under the transformation of the metric tensor $g_{\mu \nu} (x) \rightarrow \Omega^{2}(x) g_{\mu \nu} (x)$, where  $\Omega(x)$ is a smooth strictly positive function.  Such restrictions result an unique action
\begin{eqnarray}
I_{w}=-\alpha_{g} \int d^{4}x \sqrt{-g} C_{\lambda\mu\nu\kappa} C^{\lambda\mu\nu\kappa},
\end{eqnarray}
where $\alpha_{g}$ is a dimensionless coupling constant and $C_{\lambda\mu\nu\kappa}$ is the Weyl tensor \cite{weyl1918}. The action leads to a fourth order field equation \cite{weylrot5}:
\begin{eqnarray}
4 \alpha_{g} W^{\mu\nu} = 4 \alpha_{g} (C_{;\lambda;\kappa}^{\lambda\mu\nu\kappa} - \frac{1}{2} R_{\lambda\kappa} C^{\lambda\mu\nu\kappa} ) = T^{\mu\nu} , 
\end{eqnarray}
where $T^{\mu\nu}$ is the matter-energy tensor and   `;' denotes covariant derivative. It has been shown that, for a spherically symmetric mass distribution,  the resultant acceleration of a test particle takes the following form \cite{weylcluster2,weylcluster1}:
\begin{align}
&
\begin{aligned}[t]
&a(r)  \\
&= G \left[-{I_0(r)\over r^2} + {1\over R_0^2}\left({I_2(r)\over 3 r^2} - {2\over 3} r E_{-1}(r) 
- I_0(r)\right) \right] \\
& + {GM_0\over R_0^2}  - \kappa c^2 r \; .
\label{weyleq}
\end{aligned}
\end{align}
where $I_n(r)=4\pi \int_0^r \rho(x)x^{n+2} dx$ and $E_n(r)=4\pi \int_r^{+\infty} \rho(x)x^{n+2}dx$ are the interior and exterior mass moments respectively. $M_0$, $R_0$ and $\kappa$ are three conformal gravity parameters. The numerical values of the parameters have been obtained through fitting galaxy rotation curves \cite{weylrot2,weylrot1,weylrot3,weylrot4}: $R_0=24$~kpc and $M_0= 5.6\times 10^{10}M_\odot$ and $\kappa = 9.54 \times 10^{-54} $ $cm^{-2}$. These parameters are treated as universal constants in conformal gravity which implies that their values do not depend on the mass or size of the galaxy. In this work, we therefore fix $R_0$, $M_0$ and $\kappa$ to their literature values.
%%%%%%%%%%%%%%%%%%%%%%%%%%%%%%%%%%%%%%%%%%%%%%%%%%%%%%%%%%%%%%%%%%%%%%%%%%%%%%%%%%%%%%%%%%%%%%%%%%%%%%%%%%%%%%%%%%%%%%%%%%%%%%%%%%%%%%%%%%%%%%%%%%%%%%%%%%%%%%%%%%%%%%%%%%%%%%%%%%%%%%%%%%%
\subsection{MOND}
Modified Newtonian Dynamcies (MOND) \cite{mond1,famaey2012modified} is a phenomenological theory of gravity which tries to explain the observed galactic rotation curves by modifying the Newtonian acceleration law. The modification is achieved by introducing an interpolating function $\mu$ which projects the Newtonian acceleration $a_N$ to the MOND acceleration $a$ in the following way:
\begin{equation}
\mu \left(\frac{a}{a_0}\right) a = a_{N}.
\end{equation}
The interpolating function $\mu(x) \approx x$ when $x \ll 1$ and $\mu(x) \approx 1$ when $x \gg 1$. Therefore, when the galactic gravitational field is strong, the MOND acceleration exhibits Newtonian behavior. The quantity $a_0$ is the critical value below which MOND deviates from Newtonian gravity. The exact functional form of the interpolating function $\mu(x=\frac{a}{a_0})$ is not fixed. In this paper, we would use the`standard' form: 
\begin{equation}
\mu (x) = \frac{x}{\sqrt{(1 + x^{2})}},
\end{equation}
where $a_0$ = $1.14 \times 10^{-10} m/s^{2}$. The MOND acceleration reads \cite{mond1}
\begin{equation}
a_{MOND} =  \frac{a_{N}}{\sqrt{2}}\left( 1 +  \Big( 1 + \left(2a_0/a_{N}\right)^2 \Big)^{1/2} \right)^{1/2},
\end{equation}
where $a_{N}=\frac{GM(r)}{r^{2}}$ is the Newtonian acceleration generated through the baryonic mass.
\begin{figure*}[ht]
	\centering
	\includegraphics[width=0.85\linewidth]{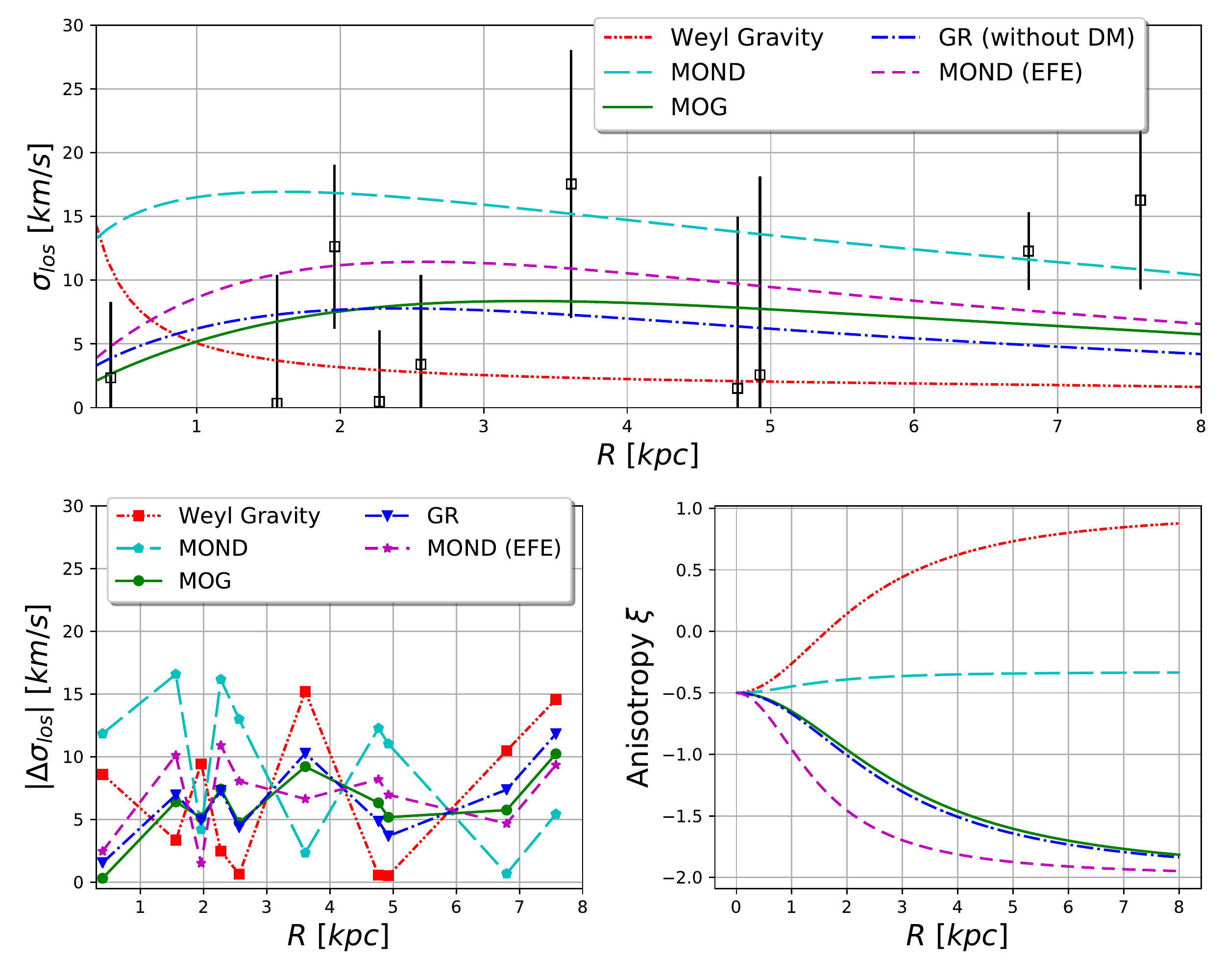}
	\caption[]{\textbf{Line-of-sight velocity dispersion profile of NGC1052-DF2 at $D=20$ Mpc}: \textit{Upper Panel :} The black squares with error-bars denote the individual dispersion measurement. On top of these, we plot the best-fit Weyl gravity dispersion profile in dash dot dotted red line, MOND profile in long dashed cyan line, MOND with EFE in short dashed magenta line and MOG profile in  solid green line. For the GR profile (long dash dotted blue line), we assume no dark matter. \textit{Lower Panel :} The residuals are plotted in the left and the best-fit anisotropy profiles are shown in the right. Details are in text.}
	\label{fig1}
\end{figure*}

The formulation presented above is strictly for the isolated MOND case for which the internal acceleration of the system $a$ is much larger than the external field $a_{ext}$ i.e. $a \gg a_{ext}$. When the external field becomes dominant ($a \ll a_{ext}$), one can safely ignore the internal field. However, when they are comparable, both must be taken into account. This particular effect, unique to MOND, is known as the External Field Effect (EFE) \cite{famaey2018mond,haghi2019new}. To capture EFE appropriately, we modify the MOND acceleration law as:
\begin{equation}
(a+a_{ext}) \mu(\frac{a+a_{ext}}{a_0})=a_{N}+a_{ext}\mu(\frac{a_{ext}}{a_0}).
\end{equation}
One can then compute $a$ if $a_{ext}$ is known. In this work, we fix the value of $a_0$ to its literature value of $1.14 \times 10^{-10} m/s^{2}$ whereas $a_{ext}$ is considered as a free parameter. Its value is computed in a case-by-case basis (Details are in Section \ref{sec4} and \ref{sec5}).
%%%%%%%%%%%%%%%%%%%%%%%%%%%%%%%%%%%%%%%%%%%%%%%%%%%%%%%%%%%%%%%%%%%%%%%%%%%%%%%%%%%%%%%%%%%%%%%%%%%%%%%%%%%%%%%%%%%%%%%%%%%%%%%%%%%%%%%%%%%%%%%%%%%%%%%%%%%%%%%%%%%%%%%%%%%%%%%%%%%%%%%%%%%
\subsection{MOG}
Scalar-Tensor-Vector Gravity (STVG) or the Modified Gravitational (MOG) theory is a covariant scalar-tensor-vector gravity theory which features a massive vector field $\phi_{\mu}$ and three scalar fields $G$, $\mu$ and $\omega$ \cite{mog}. Instead of the Newtonian gravitational constant $G_N$, the theory allows a dynamical gravitational parameter $G$ which varies with space and time. In this theory the acceleration within an arbitrary spherically symmetric mass distribution takes the following form:
\begin{align}
\begin{aligned}
a_{MOG}=&- \int_{0}^{r}    dr'\frac{2 \pi G_N r'}{\mu r^2}\rho(r')\Big\{2(1+\alpha) \\
& + \alpha(1+\mu r)[e^{-\mu(r+r')}-e^{-\mu(r-r')}] \Big\}\\
& -\int_{r}^{\infty}    dr'\frac{2 \pi G_N r'}{\mu r^2}\rho(r')\alpha\\ 
&\times \Big\{ (1+\mu r)[e^{-\mu(r+r')}-(1-\mu r)e^{-\mu(r'-r)}]\Big\}
\end{aligned}
\end{align}
where $\alpha$ and $\mu$ controls the strength and range of the attractive force, and $G_N$ is the Newtonian gravitational constant. The parameter $\alpha$ and $\mu$ depends on the mass scale of the system. The value of $\alpha$ is obtained from the relation
\begin{equation}
\alpha = \alpha_{\inf} \frac{M}{(\sqrt{M}+E)^2};
\label{alpha}
\end{equation}
and $\mu$ is given by:
\begin{equation}
\mu = \frac{D}{\sqrt{M}},
\label{mu}
\end{equation}
with $\alpha_{\inf}=10$, $D=6.25\times 10^3$ $M_{\odot}^{1/2} $ $kpc^{-1}$ and $E=2.5\times 10^4$ $M_{\odot}^{1/2} $ where $\alpha_{\inf}$ related to the effective gravitational field at infinity and the constants $D$ and $E$ has been estimated from astrophysical data \cite{moffat2008testing,moffat2009fundamental}. We treat $\alpha_{\inf}$, $D$ and $E$ to be constant in this work and set them to their literature values. This implies that $\alpha$ and $\mu$ would take different values in different galaxies.
%%%%%%%%%%%%%%%%%%%%%%%%%%%%%%%%%%%%%%%%%%%%%%%%%%%%%%%%%%%%%%%%%%%%%%%%%%%%%%%%%%%%%%%%%%%%%%%%%%%%%%%%%%%%%%%%%%%%%%%%%%%%%%%%%%%%%%%%%%%%%%%%%%%%%%%%%%%%%%%%%%%%%%%%%%%%%%%%%%%%%%%%%%%%%%%%%%%%%%%%%%%%%%%%%%%%%%%%%%%%%%%%%%%%%%%%%%%%%%%%%%%%%%%%%%%%%%%%%%%%%%%%%%%%%%%%%%%%%%%%%%%%

%%%%%%%%%%%%%%%%%%%%%%%%%%%%%%%%%%%%%%%%%%%%%%%%%%%%%%%%%%%%%%%%%%%%%%%%%%%%%%%%%%%%%%%%%%%%%%%%%%%%%%%%%%%%%%%%%%%%%%%%%%%%%%%%%%%%%%%%%%%%%%%%%%%%%%%%%%%%%%%%%%%%%%%%%%%%%%%%%%%%%%%%%%%%%%%%%%%%%%%%%%%%%%%%%%%%%%%%%%%%%%%%%%%%%%%%%%%%%%%%%%%%%%%%%%%%%%%%%%%%%%%%%%%%%%%%%%%%%%%%%%%%
\begin{figure*}[ht]
	\centering
	\includegraphics[width=0.85\linewidth]{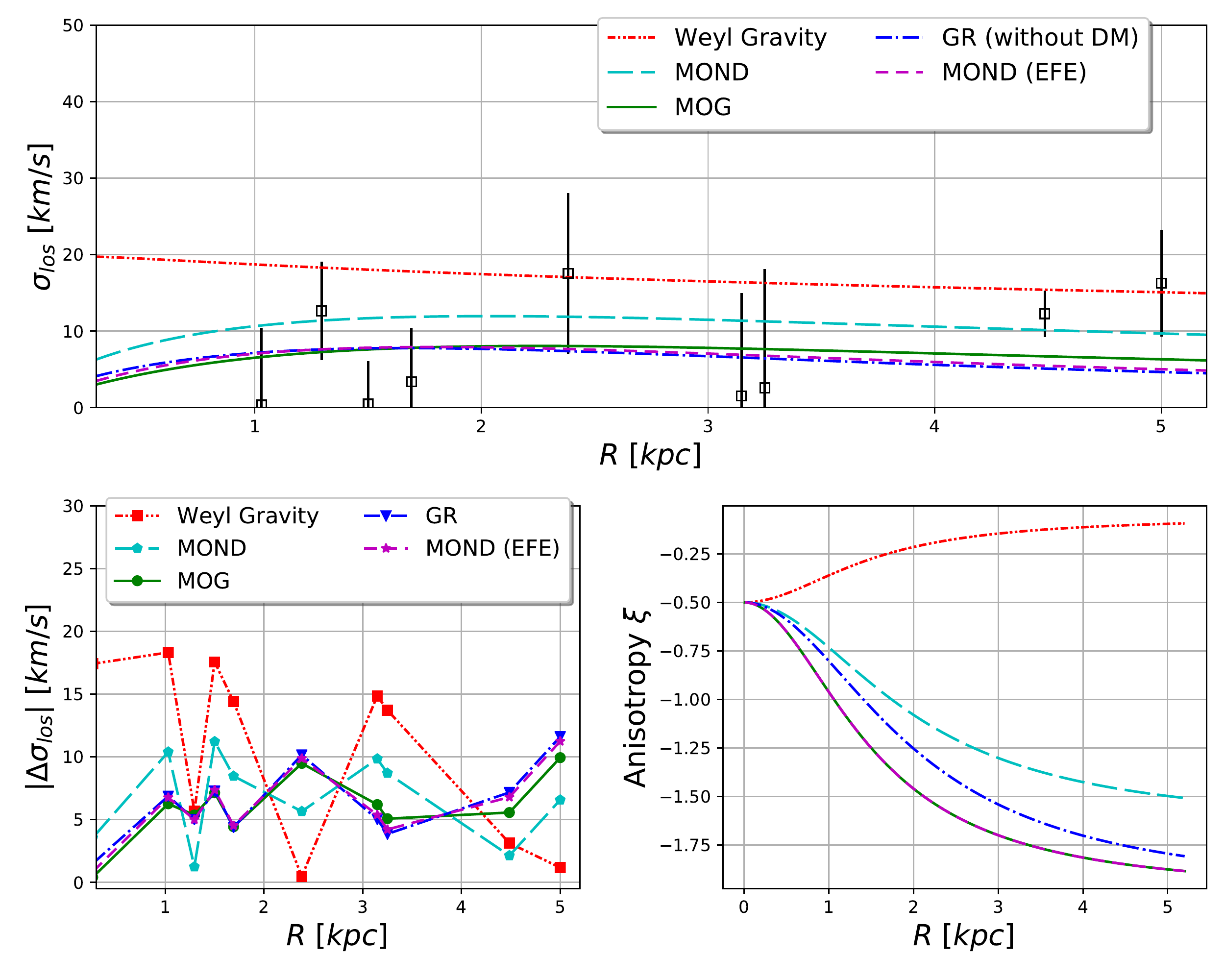}
	\caption[]{\textbf{Line-of-sight velocity dispersion profile of NGC1052-DF2 at $D=13.2$ Mpc}: \textit{Upper Panel :} The black squares with error-bars denote the individual dispersion measurement. On top of these, we plot the best-fit Weyl gravity dispersion profile in dash dot dotted red line, MOND profile in long dashed cyan line, MOND with EFE in short dashed magenta line and MOG profile in  solid green line. For the GR profile (long dash dotted blue line), we assume no dark matter. \textit{Lower Panel :} The residuals are plotted in the left and the best-fit anisotropy profiles are shown in the right. Details are in Section \ref{sec4} of text.}
	\label{fig2}
\end{figure*}

\section{Jeans Analysis and Velocity Dispersion}
\label{sec3}
We use the Jeans equation  to estimate the velocity dispersion profile of a non-rotating system with spherically symmetric mass distribution (ultra-diffuse galaxies in our case) \cite{bt1987}. The equation reads:
\begin{equation}
\frac{\partial(\rho(r)\sigma_{r}^2(r))}{\partial r} +\frac{2 \rho(r) \xi(r) \sigma_{r}^{2}(r)}{r} = \rho(r) a(r),
\label{eqn15}
\end{equation}
where $r$ is the radial distance from the center of the object, $\rho(r)$ is the radial density function and $\xi(r)$ is the anisotropy parameter. The anisotropy parameter $\xi(r)$ is defined as:
\begin{equation}
\xi(r)=1-\frac{\sigma_{\theta}^2}{\sigma_{r}^{2}},
\end{equation}
with $\sigma_{\theta}$ and $\sigma_{r}$ being the azimuthal and radial  velocity dispersions respectively. In general, $\xi(r)$ is a function of $r$. As $\lim\limits_{r\rightarrow\infty}\rho(r)=0$, we get  $\lim\limits_{r\rightarrow\infty}\rho(r)\sigma_{r}^2(r)=0$. This allows us to write:
\begin{equation}
\sigma_{r}^2(r)=\frac{1}{\rho(r)}\int\limits_r^\infty\rho(r')a(r')(\frac{r'}{r})^{2\xi(r)}~dr'.
\label{eqn16}
\end{equation}
The projected line-of-sight (LOS) velocity dispersion can then be written as:
\begin{equation}
\sigma_\mathrm{LOS}^2(R)=\frac{\int_R^\infty\frac{ r\sigma_{r}^2(r)\rho(r)}{\sqrt{r^2-R^2}}~dr - R^{2}\int_R^\infty\frac{ \xi(r) \sigma_{r}^2(r)\rho(r)}{r\sqrt{r^2-R^2}}~dr}{\int_R^\infty r\rho(r)/\sqrt{r^2-R^2}~dr},
\label{eqn17}
\end{equation}
where $R$ is the projected distance from the center of the object.\\

In reality, anisotropy is known to very with distances from the the center of the galaxy. We therefore consider generalized Osipkov-Merritt anisotropy model \cite{osipkov1979pis,merritt1985relaxation} based on three parameters (inner anisotropy, outer anisotropy, and transition radius):
\begin{equation}
\xi_{const}(r)=\xi_0 + (\xi_\infty-\xi_0)\frac{r^2}{r^2+a_\xi^2},
\end{equation}
where $\xi_0$, $\xi_\infty$ and $a_\xi$ denote the inner anisotropy, outer anisotropy, and transition radius respectively. Generalized Osipkov-Merritt anisotropy model reduces to a constant anisotrpy model when $\xi_0=\xi_\infty$.

%%%%%%%%%%%%%%%%%%%%%%%%%%%%%%%%%%%%%%%%%%%%%%%%%%%%%%%%%%%%%%%%%%%%%%%%%%%%%%%%%%%%%%%%%%%%%
\begin{table}[h]
	\centering
	\caption[]{\textbf{DF2: Best-fit parameter values and reduced chi-square values as goodness-of-fits for different theories of gravity}. $D=20.0$ $Mpc$ and no dark matter are assumed.}
	\label{T1}
	\begin{tabular}{c c c c c c}
		\hline 
		\hline
		&$\gamma$ &$\xi_0$ &$\xi_\infty$ &$a_\xi$ &$\chi^{2}/dof$ \\
		\hline
		MOG &1.23 &-0.5 &-2.00 &3.00 & 1.67\\
		MOND &1.00 &-0.5 &-0.32 &1.50 &3.47\\
		MOND with EFE &1.00 &-0.5 &1.50 &-1.99 &1.90\\
		Weyl Gravity &1.00 &-0.5 &0.99 &2.29 &3.80\\
		GR (without DM) &1.36 &-0.5 &-2.00 &2.80 &2.18\\
		\hline 
		\hline
	\end{tabular} 
\end{table}
%%%%%%%%%%%%%%%%%%%%%%%%%%%%%%%%%%%%%%%%%%%%%%%%%%%%%%%%%%%%%%%%%%%%%%%%%%%%%%%%%%%%%%%%%%%%%
\begin{figure*}[ht]
	\centering
	\includegraphics[width=0.95\linewidth]{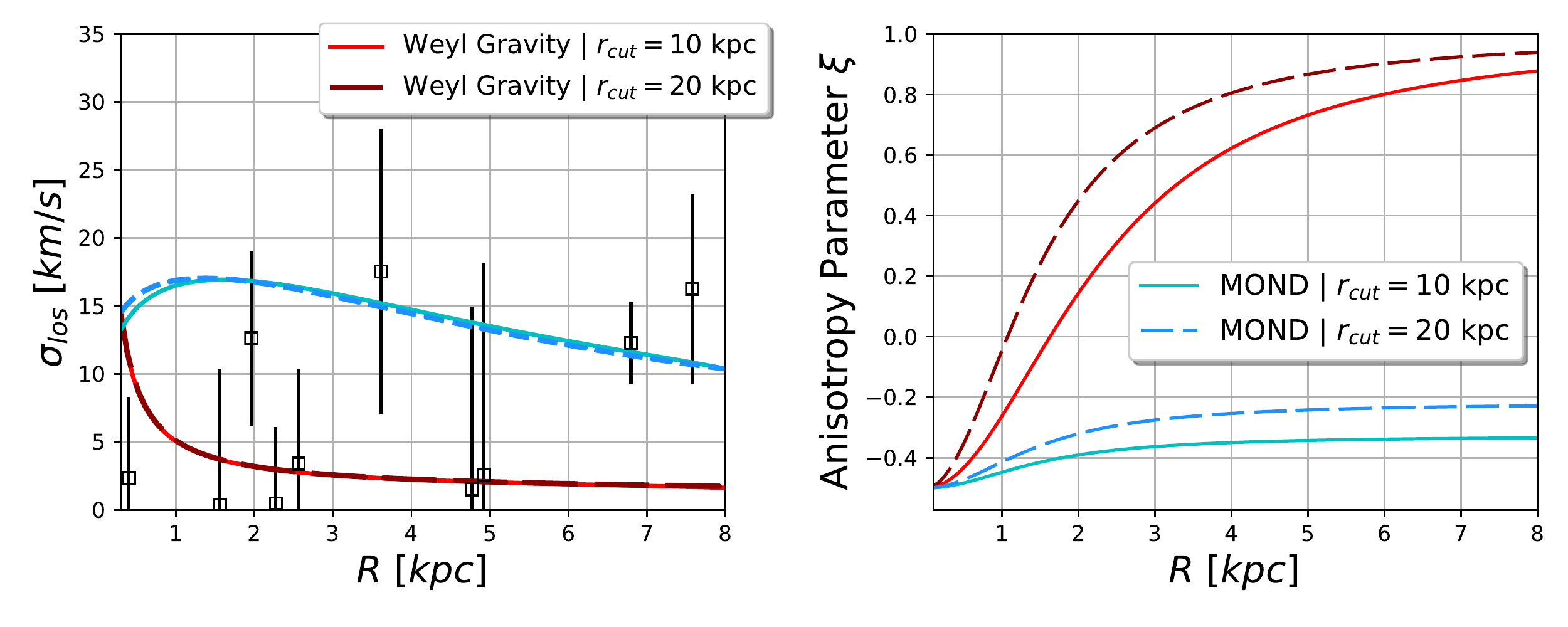}
	\caption{\textit{Left Panel :} Best-fit MOND and Weyl Gravity profiles for two different values of phenomenologically chosen parameter $r_{cut}$ ($=[10,20]$ kpc]) for NGC 1052-DF2. \textit{Right Panel :} Corresponding best-fit anisotropy profiles. Color codes are given in the legend. Details are in Section \ref{sec4b}.}
	\label{fig3}
\end{figure*}

%%%%%%%%%%%%%%%%%%%%%%%%%%%%%%%%%%%%%%%%%%%%%%%%%%%%%%%%%%%%%%%%%%%%%%%%%%%%%%%%%%%%%%%%%%%%%%%%%%%%%%%%%%%%%%%%%%%%%%%%%%%%%%%%%%%%%%%%%%%%%%%%%%%%%%%%%%%%%%%%%%%%%%%%%%%%%%%%%%%%%%%%%%%
\section{NGC 1052-DF2 : a galaxy with extermely low velocity dispersion}
\label{sec4}
\subsection{Mass Profile}
 The surface brightness of the ultra-diffuse galaxy NGC1052-DF2 can be modelled using a two-dimensional S\'{e}rsic profile, with index $n = 0.6$, effective radius $R_e = 2.2$ kpc and total luminosity $L=1.2\times10^{8}$ $L_{\odot}$ \cite{van2018galaxy}. This indicates a total mass of $M\sim2.0\times 10^{8}M_{\odot}$ within a radius $7.6$ kpc (assuming standard stellar population modelling). The resultant mass density can be closely approximated as \cite{moffat2018ngc}: 
\begin{equation}
\rho_{ser}=\gamma \frac{20 \varSigma_0}{63 R_e} exp\Big[   -\left(\frac{11r}{10R_e}\right)^{4/3} \Big],
\label{mass}
\end{equation}
where $\gamma$ is the mass-to-light ratio, $\varSigma_0=1.25 \times 10^7 M_{\odot}/kpc^2$ is the characteristic surface mass density and $R_e=2.0$ $kpc$ is the effective radius.

\subsection{Modified Gravity Fits}
\label{sec4b}
\paragraph{Assumptions :} For the anisotropy profile, generally it is assumed that the transition radius $a_\xi$ matches the effective radius of the tracer distribution ($R_e$ for our case). This is called Tied Anisotropy Number Density assumption. In our fitting exercise, we, however, do not fix $a_\xi=R_e$. Rather, we allow $a_\xi$ to take values around the value of the effective radius $R_e$. For each theory of gravity, we treat the corresponding gravitational parameters as constants. Only free parameters are: dynamical mass-light ratio $\gamma$ and three anisotropy parameters $\xi_0$, $\xi_\infty$ and $a_\xi$.  

\paragraph{Best-fit profiles with D=20 Mpc :} We assume a distance (from us) $D=20$ $Mpc$ for NGC 1052-DF2 \cite{van2018galaxy,van2018revised,van2019second} and allow the mass-light ratio and  anisotropy parameters to vary. Additionally, we impose a sharp trimming of the mass profile at $r_{cut}=10$ kpc to model tidal stripping of DF2 due to nearby massive galaxies (keeping in mind that the radial distance of the last observed GC is $\sim$8 kpc). In MOND, within the context of EFE, the external acceleration due to the host NGC 1052 (residing at a distance of 80-90 kpc) is $a_{ext}\sim$0.15$a_0$ \cite{famaey2018mond}. We, therefore, set $a_{ext}=0.15 a_0$ to compute the effective MOND profile with EFE. 

We first present a comparison between the data \cite{wasserman2018deficit} and the best-fit line-of-sight (los) velocity dispersion profiles in modified gravity theories in Figure \ref{fig1}. The individual GC velocity dispersion measurements are shown in black squares along with their quoted error-bars. The resultant best-fit modified gravity dispersion profiles from the inferred baryonic mass (long dash dotted blue line for GR; dash dot dotted red line for Weyl gravity; long dashed cyan line for MOND; short dashed magenta line for MOND with EFE and solid green line for MOG) are then superimposed in Figure \ref{fig1} (upper panel). The absolute residuals and best-fit anisotropy profiles are plotted in the lower panel of Figure \ref{fig1}. The best-fit parameter values and the corresponding reduced chi-square values are reported in Table \ref{T1}.
\begin{figure*}[hbt]
	\centering
	\includegraphics[width=0.75\linewidth]{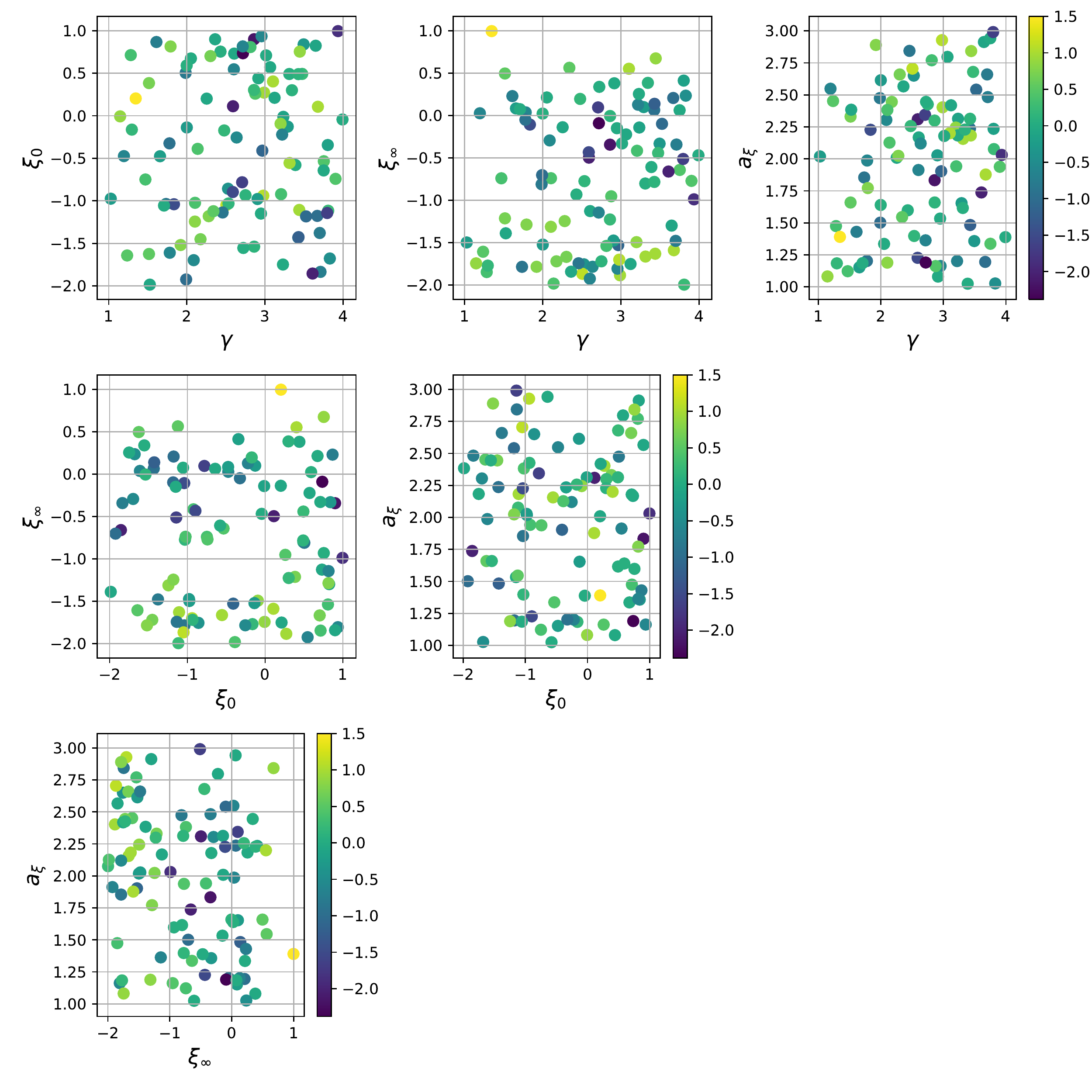}
	\caption{`Likelihood' (i.e. log of chi-square) between best-fit MOND dispersion profile for NGC 1052-DF2 and the dispersion profiles yielded by randomly chosen model parameters as a function of different model parameters; x-axis and y-axis denotes respective slice of model parameters while `Likelihood' is denoted by colormap. (\textit{Section \ref{sec4c} in text)}}
	\label{fig4}
\end{figure*}
%%%%%%%%%%%%%%%%%%%%%%%%%%%%%%%%%%%%%%%%%%%%%%%%%%%%%%%%%%%%%%%%%%%%%%%%%%%%%%%%%%%%%%%%%%%%%
\begin{table}[h]
	\centering
	\caption[]{\textbf{DF2: Best-fit parameter values and reduced chi-square values as goodness-of-fits for different theories of gravity}. $D=13.2$ $Mpc$ and no dark matter are assumed.}
	\label{T2}
	\begin{tabular}{c c c c c c}
		\hline 
		\hline
		&$\gamma$ &$\xi_0$ &$\xi_\infty$ &$a_\xi$ &$\chi^{2}/dof$ \\
		\hline
		MOG &1.88 &-0.5 &-2.00 &1.47 & 1.61\\
		MOND &1.00 &-0.5 &-0.72 &1.01 &2.31\\
		MOND with EFE &1.00 &-0.5 &-2.00 &-1.01 &1.81\\
		Weyl Gravity &1.00 &-0.5 &-0.06 &2.00 &4.95\\
		GR (without DM) &1.87 &-0.5 &-2.00 &1.00 &2.11\\
		\hline 
		\hline
	\end{tabular} 
\end{table}
%%%%%%%%%%%%%%%%%%%%%%%%%%%%%%%%%%%%%%%%%%%%%%%%%%%%%%%%%%%%%%%%%%%%%%%%%%%%%%%%%%%%%%%%%%%%%

We find that all of the theories in question favors a tangential anisotropy ($\xi < 0$) in the interior of the galaxy. For Weyl gravity and MOND, the best-fit anisotropy profiles become more radially anisotropic in the outer region of the galaxy while MOG, MOND with EFE and GR (without DM) favour a model that becomes more tangentially anisotropic as distances grow. The later two theories exhibit a higher mismatch in the interior of the DF2. Though the chi-square values of Weyl gravity and MOND are slightly higher than GR and MOG, we find that all five models can reasonably fit the dispersion data with acceptable mass-to-light ratio. It is important to note that the galaxy is considered to be devoid of `dark matter' in the $\Lambda$$CDM$ context. Therefore, GR (without DM) should be able to fit the data. This is the reason we also show best-fit GR (without DM) profile in Figure \ref{fig1}.

\paragraph{Best-fit profiles with D=13.2 Mpc :} We now consider the possibility that the distance of NGC 1052-DF2 is much smaller than 20 Mpc. Trujillo \textit{et. al} \cite{trujillo2019distance} reports an estimation of $D\sim13.2$ Mpc. Our previous work \cite{islam2019modified} shows that the agreement between data and modified gravity models  is better  for  such  smaller  distance  estimates. However, the  was done with different assumptions, and is not clear whether this will hold true with more generalized assumptions made in this work. We, therefore, have redone our analysis with distance $D=13.2$ Mpc. The galactocentric distances of all GCs have been rescaled to $D=13.2$ Mpc. Total mass and effective radius of NGC 1052-DF2 at this distances are estimated to be around $\sim 6\times10^7$ $M_{\odot}$ and $R_e=1.4 \pm 0.1$ kpc respectively \cite{trujillo2019distance}. At this distance, the last observed GC has  a distance of 5.1 kpc from the center of the galaxy. Keeping this in mind, we choose $r_{cut}$ to be $8.0$ kpc.

We find that overall the modified gravity fits improve a bit except for Weyl gravity which shows a relatively poor but still acceptable fit with a reduced $\chi^2$ value of 4.95 (against 3.80 for $D=20$ Mpc) (Figure \ref{fig2}). The best-fit parameter values and the corresponding reduced chi-square values are reported in Table \ref{T2}. The best-fit anisotropy profile for Weyl gravity exhibits a relatively more tangential nature than for the scenario with $D=20$ Mpc. All other theories of gravity prefer more radial nature in anisotropy.

%%%%%%%%%%%%%%%%%%%%%%%%%%%%%%%%%%%%%%%%%%%%%%%%%%%%%%%%%%%%%%%%%%%%%%%%%%%%%%%%%%%%%%%%%%%%%
\begin{table}[h]
	\centering
	\caption[]{\textbf{DF2: Reduced chi-square values as goodness-of-fits for Weyl gravity and MOND}. Two different values of trimming radius assumed: $r_{cut}=\{10,20\}$ kpc. No dark matter are assumed.}
	\label{T3}
	\begin{tabular}{c c c c c c}
		\hline 
		\hline
		&$\gamma$ &$\xi_0$ &$\xi_\infty$ &$a_\xi$ &$\chi^{2}/dof$ \\
		\hline
		MOND: $r_{cut}=10$ kpc&1.00 &-0.5 &-0.32 &1.50 &3.47\\
		MOND: $r_{cut}=20$ kpc &1.00 &-0.5 &-0.21 &1.50 & 3.62\\
		Weyl: $r_{cut}=10$ kpc&1.00 &-0.5 &0.99 &2.29 &3.80\\
		Weyl: $r_{cut}=20$ kpc &1.00 &-0.5 &-2.00 &1.50 &3.47\\
		\hline 
		\hline
	\end{tabular} 
\end{table}
%%%%%%%%%%%%%%%%%%%%%%%%%%%%%%%%%%%%%%%%%%%%%%%%%%%%%%%%%%%%%%%%%%%%%%%%%%%%%%%%%%%%%%%%%%%%%

\begin{figure*}[htb]
	\centering
	\includegraphics[width=0.85\linewidth]{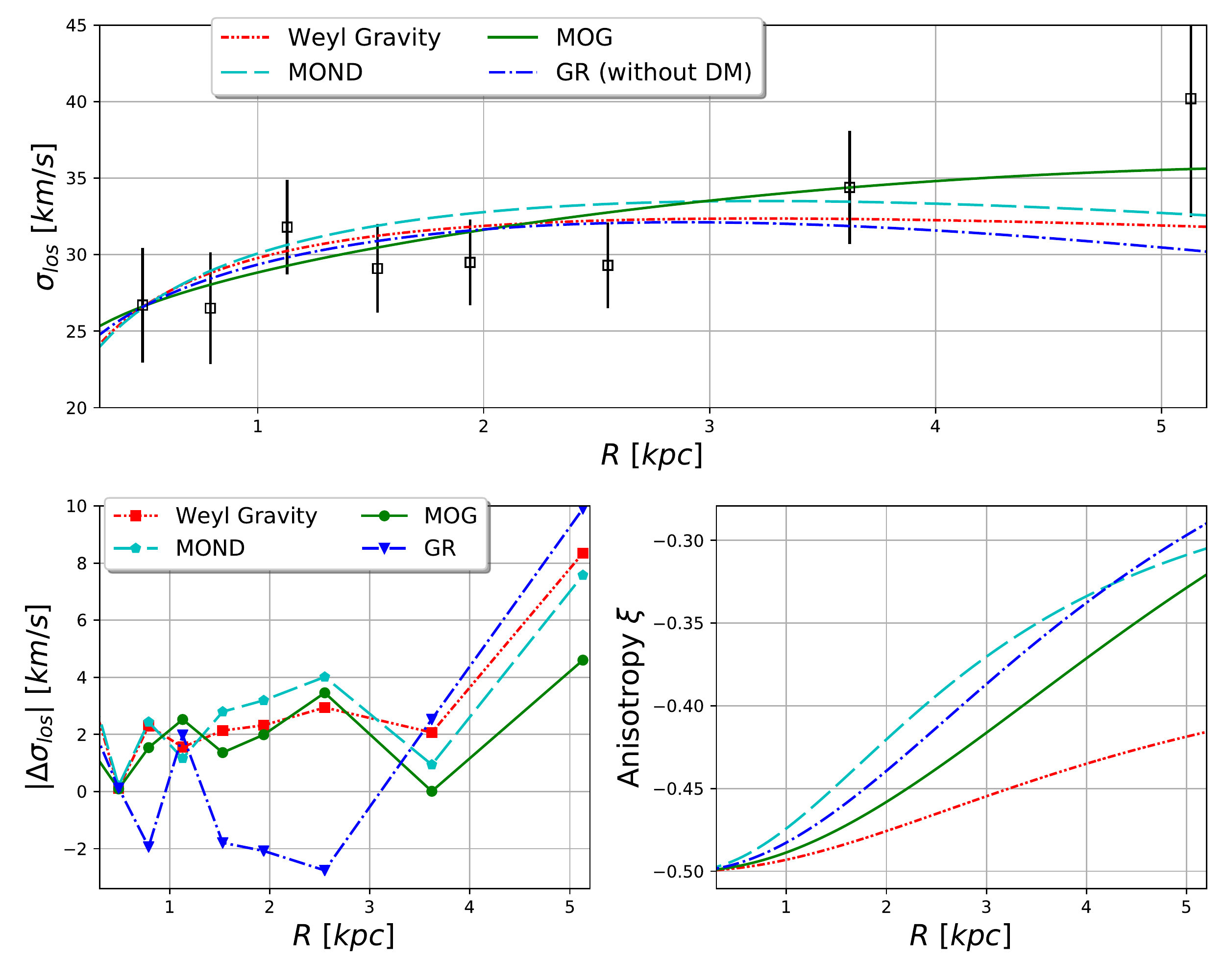}
	\caption[]{\textbf{Line-of-sight velocity dispersion profile of NGC1052-DF44 at $D=20$ Mpc}: \textit{Upper Panel :} The black squares with error-bars denote the individual dispersion measurement. On top of these, we plot the best-fit Weyl gravity dispersion profile in dash dot dotted red line, MOND profile in long dashed cyan line and MOG profile in  solid green line. For the GR profile (long dash dotted blue line), we assume no dark matter. \textit{Lower Panel :} The residuals are plotted in the left and the best-fit anisotropy profiles are shown in the right. Details are in text.}
	\label{fig5}
\end{figure*}

\paragraph{Effect of $r_{cut}$ :} We have already mentioned that the value the of trimming radius $r_{cut}$ is chosen phenomenologically in such a way that it is greater than the last observed data point and atleast thrice the effective radius of the galaxy. However, we would like to note that increasing the value of $r_{cut}$ will only slightly increase the dispersion values in the outskirts. There will be nominal changes in the result as the density itself drops rapidly (exponentially) which ensures the outer-most portions of galaxy will have little effect on the velocity dispersion. As a demonstration, we repeat the fitting exercise for Weyl gravity and MOND (assuming $D=20$ Mpc) for a different choice of $r_{cut}=20$ kpc. The best-fit parameters and chi-squre values are presented in Table \ref{T3}. We find that the best-fit velocity dispersion profiles does not change much leaving reduced $\xi^2$ values virtually unchanged (Figure \ref{fig3}). We, however, notice a change in the best-fit anisotropy profiles for Weyl gravity and MOND. This hints a possible degeneracy between dynamical mass-light ratio $\gamma$ and anisotropy parameters $\xi_0$, $\xi_\infty$ and $a_\xi$.

\subsection{Possible Degeneracies in Model}
\label{sec4c}
To investigate the possible degeneracies between $\gamma$, $\xi_0$, $\xi_\infty$ and $a_\xi$, we fix the theory of gravity. For the demonstration, we choose MOND as the description of gravitational field. We then randomly sample $\gamma$ from $[1,4]$, $\xi_0$ from $[-2,1]$, $\xi_\infty$ from $[-2,1]$ and $a_\xi$ from $[1,3]$ for NGC 1052-DF2. Distance is assumed to be $D=20$ Mpc. We then measure the `Likelihood' between the MOND profile generated by the sampled \{$\gamma$, $\xi_0$, $\xi_\infty$, $a_\xi$\} and the best-fit MOND profile showed in Figure \ref{fig1}. The `Likelihood' is quantified by the log of chi-square values between these two MOND profiles. Smaller value of `Likelihood' indicates a better match between these two profiles. We then plot the Likelihood using a heat-map in Figure \ref{fig4} as a function of the model parameters. We observe that different combinations of parameter values yield similar `Likelihood' values implying underlying degeneracies between these parameters. Similar degeneracies are also expected in other theories of gravity.

\section{NGC 1052-DF44 : galaxy with extremely high velocity dispersion}
\label{sec5}
DF44 is very similar to DF2 in terms of size and luminosity. However, it shows an extraordinary high velocity dispersions of $\sigma \sim 41.0 \pm 8$ km/s \cite{Dokkum_2016}. For DF44, we assume a mass profile similar to DF2 (Eq. \ref{mass}) but with the characteristic surface mass density $\varSigma_0=1.372 \times 10^7 M_{\odot}/kpc^2$ and effective radius $R_e=4.6$ $kpc$. For this galaxy, we use $r_{cut}=20$ kpc. Though phenomenological, this choice is still reasonable as the $r_{cut}=20$ kpc is way larger than the distance at which last data GC has been observed for DF44 (5.1 kpc) and there is no evidence of nearby massive perturber. This particular galaxy is considered to be a dark matter dominated galaxy with little baryonic contents in the context of $\Lambda$$CDM$. Thus, only GR (without DM) model with reasonable stellar mass-to-light ratio would not be able to fit the data. The best-fit mass-to-light ratio would be absurdly high. On the other hand, a modified gravity theory that do not invoke the existence of dark matter should be able to fit the dispersion data with mass-to-light ratio values of the order of unity. As there is no report of nearby massive galaxies for DF44, it is unlikely that DF44 is subjected to any external gravitational pull. Therefore, we do not consider MOND under EFE. In Figure \ref{fig5}, we show the best-fit modified gravity dispersion profiles as well as GR profiles along with observed values. The best-fit dynamical mass-light ratio $\gamma$ and anisotropy parameters $\xi_0$, $\xi_\infty$ and $a_\xi$ in Weyl gravity, MOND, MOG and GR are presented in Table \ref{T4}.

%%%%%%%%%%%%%%%%%%%%%%%%%%%%%%%%%%%%%%%%%%%%%%%%%%%%%%%%%%%%%%%%%%%%%%%%%%%%%%%%%%%%%%%%%%%%%
\begin{table}[h]
	\centering
	\caption[]{\textbf{DF4: best-fit parameter values and reduced chi-square values as goodness-of-fits for different theories of gravity}. No dark matter are assumed.}
	\label{T4}
	\begin{tabular}{c c c c c c}
		\hline 
		\hline
		&$\gamma$ &$\xi_0$ &$\xi_\infty$ &$a_\xi$ &$\chi^{2}/dof$ \\
		\hline
		MOG &10.06 &-0.50 &-0.08 &5.98 & 0.71\\
		MOND &2.61 &-0.50 &-0.24 &3.01 &1.33\\
		Weyl Gravity &1.49 &-0.50 &-0.36 &4.50 &1.02\\
		GR (without DM) &10.50 &-0.50 &-0.13 &4.50 &1.01\\
		\hline 
		\hline
	\end{tabular} 
\end{table}
%%%%%%%%%%%%%%%%%%%%%%%%%%%%%%%%%%%%%%%%%%%%%%%%%%%%%%%%%%%%%%%%%%%%%%%%%%%%%%%%%%%%%%%%%%%%%

We observe that the best-fit mass-to-light ratio $\gamma$ for GR ($\gamma_{GR}=10.50$) is too high indicating the need for dark matter. Interestingly, MOG ($\gamma_{MOG}=10.06$) too requires unacceptably high mass-to-light ratio to fit the observed dispersion profile. Thus, we conclude that MOG is not consistent with DF44 even though it can fit DF2 very well. On the other hand, best-fit mass-to-light ratio for Weyl gravity (($\gamma_{Weyl}=1.49$) and MOND ($\gamma_{MOND}=2.61$) are acceptable for UDGs and matches with the estimates from population synthesis and stellar evolution models \cite{haghi2019star}. \\

The anomaly of DF2 and DF44 together sets a challenge for MOG. Weyl gravity and MOND, however, are found to be able to account for both the enigmatic galaxies (DF2 and DF44). For DF44, we do not take EFE into consideration as no massive nearby galaxy, that might generate an external field for DF44, has been reported.

\section{Do UDGs follow Radial Acceleration Relation (RAR) ?}
\label{sec6}

Analysis of 153 spiral galaxies by McGaugh \textit{et al} \citep{McGaugh} revealed tighter correlation between the radial acceleration inferred from the rotation curves of galaxies and the same expected from Newtonian (centripetal) acceleration generated by the baryons. This empirical relation, known as Radial Acceleration Relation (RAR), is given by:
\begin{equation}
\Huge{ a_{RAR}=\frac{a_{n}}{1-exp(-(\frac{a_{n}}{a{\dagger}})^{1/2})}, } \normalsize
\label{rar}
\end{equation}
where $a_{n}$ is the Newtonian acceleration produced by the baryonic mass only and $a{\dagger}=1.2 \times 10^{-10}$ $m s^{-2}$ is the acceleration scale. Lelli \textit{et al} \cite{lelli2017one} have further found that similar relation holds for other types of galaxies such as ellipticals, lenticulars, and dwarf spheroidals. The universality of RAR across different types of galaxies is considered to be an indication of an underlying departure from GR at the galactic scale. In this section, we would like to find out whether the RAR scaling holds true for DF2 and DF44 too.

\begin{figure}[ht]
	\centering
	\includegraphics[width=0.95\linewidth]{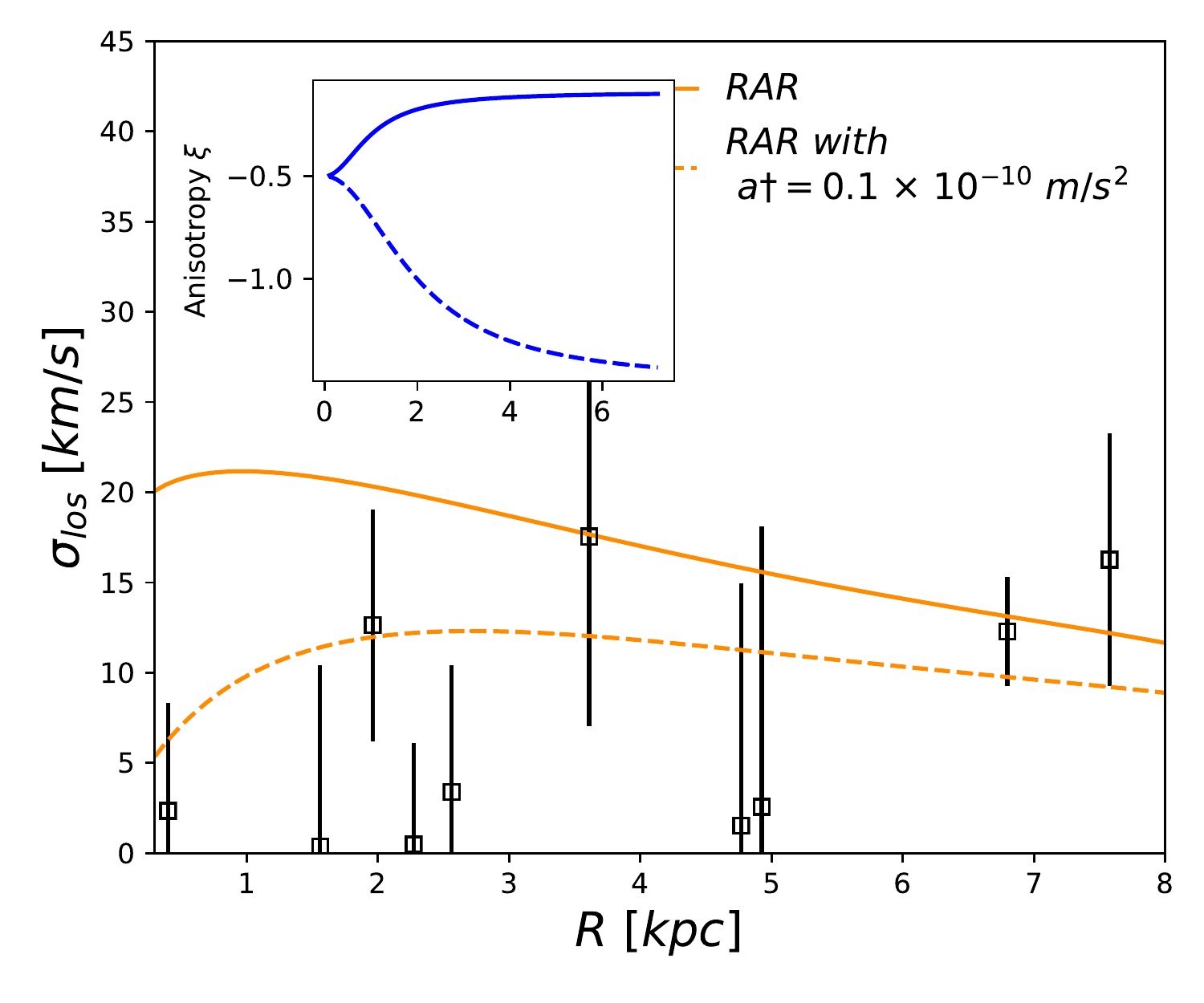}
	\caption[]{\textbf{Line-of-sight velocity dispersion profile of NGC1052-DF2 and RAR}: The black squares with error-bars denote the individual dispersion measurements. On top of these, we plot the best-fit RAR dispersion profile in orange solid line. Dashed line represent the best-fit profile if acceleration scale $a{\dagger}$ is set to a smaller value of $0.1 \times 10^{-10}$ $m s^{-2}$. The corresponding best-fit anisotropy profile for RAR with $a{\dagger}$ set to its literature value and RAR with  $a{\dagger}=0.1 \times 10^{-10}$ $m s^{-2}$ are shown in the inset in blue solid and blue dashed line respectively. Details are in Section \ref{sec6}.}
	\label{fig6}
\end{figure}

To investigate this direction, we first replace the acceleration law by Eq. \ref{rar} and compute the best-fit RAR velocity dispersion profiles for DF2 and DF44. In the case of DF2, we find RAR is unable to provide any good fit to the data unless either it is allowed to assume a dynamical mass-light ratio $\gamma_{RAR} \ll 1$ indicating a conflict with observation or the RAR scaling parameter $a{\dagger}$ is set to be at-least one order of magnitude smaller than the literature value. In Figure \ref{fig6}, we plot the resultant dispersion profile from RAR scaling along with observed values (black square) for DF2. It is clear that RAR fails to account for the observed profile with its universal acceleration scale $a{\dagger}=1.2 \times 10^{-10}$ $m s^{-2}$ (solid orange line) if the mass-light ratio is not allowed to be smaller than unity. However, we demonstrate that a smaller value of $a{\dagger}$ ($0.1 \times 10^{-10}$ $m s^{-2}$) is able to yield a good match with data with acceptable mass-light ratio of 1.00 and reduced chi-square value 1.73 (dashed orange line).

\begin{figure}[ht]
	\centering
	\includegraphics[width=0.95\linewidth]{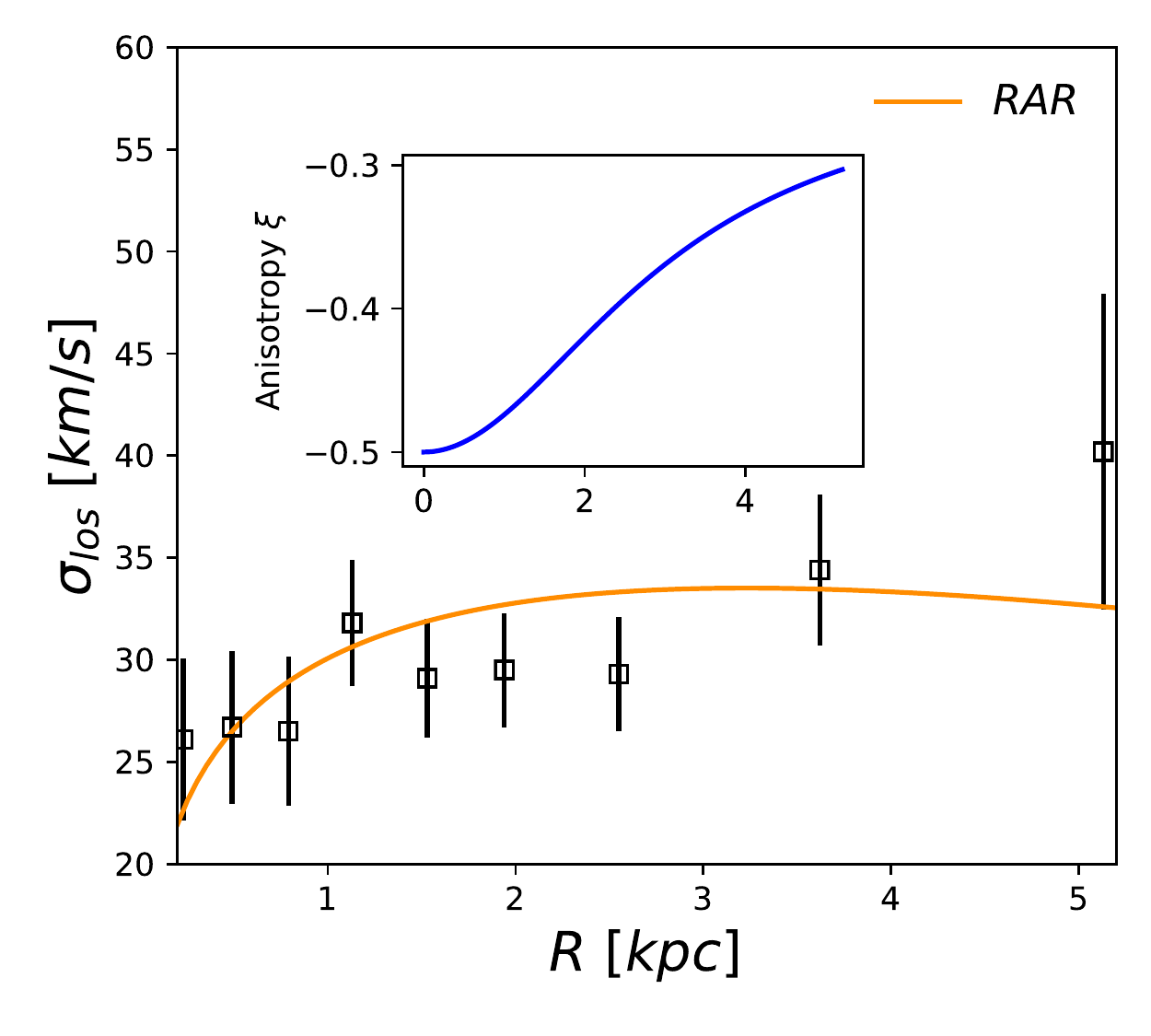}
	\caption[]{\textbf{Line-of-sight velocity dispersion profile of NGC1052-DF44 and RAR}: The black squares with error-bars denote the individual dispersion measurements. On top of these, we plot the best-fit RAR dispersion profile in orange solid line. Best-fit anisotropy profile is shown in inset. Details are in Section \ref{sec6}.}
	\label{fig7}
\end{figure}

On the other hand, RAR does an excellent job in fitting DF44 dispersion data with the original acceleration scale $a{\dagger}=1.2 \times 10^{-10}$ $m s^{-2}$ with mass-light ratio 1.13 and reduced chi-square value of 1.34. The resultant RAR profile goes through several of the DF44 dispersion data-points (Figure \ref{fig7}). 

%%%%%%%%%%%%%%%%%%%%%%%%%%%%%%%%%%%%%%%%%%%%%%%%%%%%%%%%%%%%%%%%%%%%%%%%%%%%%%%%%%%%%%%%%%%%%%%%%%%%%%%%%%%%%%%%%%%%%%%%%%%%%%%%%%%%%%%%%%%%%%%%%%%%%%%%%%%%%%%%%%%%%%%%%%%%%%%%%%%%%%%%%%%%%%%%%%%%%%%%%%%%%%%%%%%%%%%%%%%%%%%%%%%%%%%%%%%%%%%%%%%%%%%%%%%%%%%%%%%%%%%%%%%%%%%%%%%%%%%%%%%%
\section{Discussions}
\label{sec7}
In our previous paper \cite{islam2019modified} we test the modified theories of gravity (namely Weyl conformal gravity, MOND and MOG along with GR without dark matter) to explore whether it is possible to explain the dynamics of DF2 and DF4 in the context of modified gravity theories. Our analysis suggested that, for all theories in question, it is possible to fit the observed dispersion data of these UDGs with reasonable success in modified gravity paradigm. Our study assumed the anisotropy parameter to be zero throughout - an assumption which might not be true always. In this paper, we have relaxed this particular assumption and assumed a radially varying anisotropy profile.

Haghi \textit{et al.} \cite{haghi2019star} have also examined the viability of MOG and MOND using the radial dispersion data of DF44.  For MOG, they followed two different approaches. First, they fixed the value of the two MOG parameters $\mu$ and $\alpha$ to the values obtained from rotation curve fits \cite{moffat2013mog}. A fit to the data requires a high mass-to-light ratio of $\gamma=7.4$, inconsistent with stellar population synthesis models. The authors then allowed $\mu$ and $\alpha$ to vary resulting best-fit values of $\alpha=221\pm 112$ and $\mu=0.41 \pm 35$ $kpc^{-1}$. These values are significantly larger than the values obtained from galactic rotation curve surveys. We, however, took a different strategy in our analysis. Instead of treating $\mu$ and $\alpha$ to be completely independent, we use their dependences on the mass and length scale (Eq. \ref{alpha} and Eq. \ref{mu}) to infer their true values in the systems like DF2 and DF44 - an approach recommended by Moffat and Toth \cite{moffat2018ngc}. 
 
Our analysis do not explicitly incorporate any effect of tidal heating. However, we use a simplistic way to compensate the effects of tidal stripping (if any) due to nearby massive galaxies. We assume a phenomenological cut-off radius $r_{cut}$ beyond which we set the mass density to be zero. For DF2, we use $r_{cut}=10$ kpc (taking into account that the last observed GC lies about 8 kpc from the galactic center and the host (massive) galaxy NGC 1052 is only 80 kpc away). For DF44, however, we use to larger value of $r_{cut}=20$ kpc as no nearby massive galaxy has been reported. We further assume the distance of DF2 to be 20 Mpc. However, \cite{trujillo2019distance} reports a different estimation of  $D\sim13.2$ Mpc for DF2. We have already showed in our previous analysis \cite{islam2019modified} that a closer distance results a better match between data and modified gravity models. We, however, have revisited the question under current assumptions and found that the conclusion holds true.
 
Furthermore, we have explored the External Field Effects (EFE) \cite{famaey2018mond} in MOND in addition to isolated MOND computations. We find that MOND better matches the dispersion data when EFE is included. However, EFE could be safely neglected for DF44 \cite{haghi2019star} as there is no evidence of any nearby massive galaxy that can potentially exert a gravitational pull on DF44..
 
%%%%%%%%%%%%%%%%%%%%%%%%%%%%%%%%%%%%%%%%%%%%%%%%%%%%%%%%%%%%%%%%%%%%%%%%%%%%%%%%%%%%%%%%%%%%%%%%%%%%%%%%%%%%%%%%%%%%%%%%%%%%%%%%%%%%%%%%%%%%%%%%%%%%%%%%%%%%%%%%%%%%%%%%%%%%%%%%%%%%%%%%%%%%%%%%%%%%%%%%%%%%%%%%%%%%%%%%%%%%%%%%%%%%%%%%%%%%%%%%%%%%%%%%%%%%%%%%%%%%%%%%%%%%%%%%%%%%%%%%%%%%
\section{Conclusions}
\label{sec8}
In this paper, we investigated the viability of three popular modified gravity theories (Weyl conformal gravity, MOND and MOG) in the light of recent observations of anomalous velocity dispersion of NGC 1052-DF2 and NGC 1052-DF44. The first galaxy exhibits an extremely low velocity dispersion compared to its radial extent while the second galaxy has quite high dispersion values. In fact, these two galaxies mark the two extreme ends among known galaxies in terms of estimated dynamical mass (or apparent `dark matter' contents in $\Lambda$$CDM$). We find that Weyl gravity and MOND manages to fit the dispersion profiles reasonably without the need of any dark matter for both the galaxies. MOG, on the other hand, can explain the observed dispersion of DF2 easily but fails to account for DF44 unless it assumes high mass-to-light ratio (i.e., in other words, unless it assumes some amount of dark matter).

Finally, we present a preliminary investigation of whether the celebrated `Radial Acceleration Relaion' (RAR), documented in hundreds of galaxies, between expected Newtonian acceleration and observed radial acceleration holds true for DF2 and DF44. We have found that while DF44 complies with RAR with the original acceleration scale $a{\dagger}=1.2 \times 10^{-10}$ $m s^{-2}$, DF2 follows the same relation but with a different value of the acceleration scale $a{\dagger}$. In future, as more and more dispersion data for UDGs would be available, it would be exciting to statistically analyze the compatibility between RAR and UDGs.\\

%%%%%%%%%%%%%%%%%%%%%%%%%%%%%%%%%%%%%%%%%%%%%%%%%%%%%%%%%%%%%%%%%%%%%%%%%%%%%%%%%%%%%%%%%%%%%%%%%%%%%%%%%%%%%%%%%%%%%%%%%%%%%%%%%%%%%%%%%%%%%%%%%%%%%%%%%%%%%%%%%%%%%%%%%%%%%%%%%%%%%%
\noindent \textbf{Acknowledgement:}
TI thanks Koushik Dutta for fruitful discussions. This work is supported by a Doctoral Fellowship at UMass Dartmouth and NSF grant PHY180666. Computation has been carried out in the cluster CARNIE at Center for Scientific Computation and Visualization Research (CSCVR), University of Massachusetts (UMass) Dartmouth. TI thanks the referees for helpful suggestions.
%%%%%%%%%%%%%%%%%%%%%%%%%%%%%%%%%%%%%%%%%%%%%%%%%%%%%%%%%%%%%%%%%%%%%%%%%%%%%%%%%%%%%%%%%%%%%%%%%%%%%%%%
%%%%%%%%%%%%%%%%%%%%%%%%%%%%%%%%%%%%%%%%%%%%%%%%%%%%%%%%%%%%%%%%%%%%%%%%%%%%%%%%%%%%%%%%%%%%%%%%%%%%%%%%%%%%%%%%%%%%%%%%%%%%%%%%%%%%%%%%%%%%%%%%%%%%%%%%%%%%%%%%%%%%%%%%%%%%%%%%%%%%%%%%%%%%%%%%%%%%%%%%%%%%%%%%%%%%%%%%%%%%%%%%%%%%%%%%%%%%%%%%%%%%%%%%%%%%%%%%%%%%%%%%%%%%%%%%%%%%%%%%%%%%
%\bibliography{T2019UDG}

\end{document}